\documentclass[usenatbib,usegraphicx]{mn2e}
\usepackage{amsfonts}
\usepackage{amsmath}
\usepackage{amssymb}
\usepackage{comment}
\usepackage{color}

\title[Tidal dwarf galaxies in the nearby Universe]
    {Tidal dwarf galaxies in the nearby Universe}

\author[Sugata Kaviraj]
{Sugata Kaviraj$^{1,2}$\thanks{E-mail: s.kaviraj@imperial.ac.uk;
skaviraj@astro.ox.ac.uk}, Daniel Darg$^{2}$, Chris Lintott$^{2}$,
Kevin Schawinski$^{3,4}$ and \newauthor Joseph
Silk$^{2}$\\\\
$^{1}$Blackett Laboratory, Imperial College London, London SW7 2AZ,
UK\\
$^{2}$Department of Physics, University of Oxford, Keble Road,
Oxford, OX1 3RH, UK\\
$^{3}$Department of Physics, Yale University, New Haven, CT 06511,
USA\\
$^{4}$Einstein Fellow\\}

\begin{document}

\maketitle

%%%%%%%%%%%%%%%%%%%%%%%%%%%%
% symbols for references:  %
%%%%%%%%%%%%%%%%%%%%%%%%%%%%
\def \aj {AJ}
\def \mnras {MNRAS}
\def \pasp {PASP}
\def \apj {ApJ}
\def \apjs {ApJS}
\def \apjl {ApJL}
\def \aap {A\&A}
\def \nat {Nature}
\def \araa {ARAA}
\def \iaucirc {IAUC}
\def \aaps {A\&A Suppl.}
\def \qjras {QJRAS}
\def \na {New Astronomy}
\def\lesssim{\mathrel{\hbox{\rlap{\hbox{\lower4pt\hbox{$\sim$}}}\hbox{$<$}}}}
\def\gtrsim{\mathrel{\hbox{\rlap{\hbox{\lower4pt\hbox{$\sim$}}}\hbox{$>$}}}}

%.........................................................................................................

\begin{abstract}
We present a statistical observational study of the tidal dwarf
(TD) population in the nearby Universe by exploiting a large,
homogeneous catalogue of galaxy mergers compiled from the
\emph{Sloan Digital Sky Survey}. 95\% of TD-producing mergers
involve two spiral progenitors (typically both in the blue cloud),
while most remaining systems have at least one spiral progenitor.
The fraction of TD-producing mergers where both parents are
early-type galaxies is less than 2\%, suggesting that TDs are
unlikely to form in {\color{black}such} mergers. The bulk of the
TD-producing systems inhabit a field environment and have mass
ratios greater than $\sim$1:7 (the median value is 1:2.5). TDs
forming at the tidal-tail tips are $\sim$4 times more massive than
those forming at the base of the tails. TDs have stellar masses
that are less than 10\% of the stellar masses of their parents
(the median is 0.6\%) and lie within 15 optical half-light radii
of their parent galaxies. The TD population is typically bluer
than the parents, with a median offset of $\sim$0.3 mag in the
$(g-r)$ colour and the TD colours are not affected by the presence
of AGN activity in their parents. An analysis of their star
formation histories indicates that TDs contain both newly formed
stars (with a {\color{black}median} age of $\sim$30 Myrs) and old
stars drawn from the parent disks, each component probably
contributing roughly equally to the stellar mass of the object.
Thus TDs are not formed purely through gas condensation in tidal
tails but host a significant component of old stars from the
parent disks. Finally, an analysis of the TD contribution to the
observed dwarf to massive galaxy ratio in the local Universe
indicates that $\sim$6\% of dwarfs in nearby clusters may have a
tidal origin, if TD production rates in nearby mergers are
representative of those in the high-redshift Universe. Even if TD
production rates at high redshift were several factors higher, it
seems unlikely that the entire dwarf galaxy population today is a
result of merger activity over the lifetime of the Universe.

%Maybe put the fraction of gas-rich major mergers that do produce TDs here?

\end{abstract}

%.........................................................................................................


%.........................................................................................................

\section{Introduction}
Galaxy mergers are a key driver of cosmological evolution,
stimulating intense star formation episodes
\citep[e.g.][]{Barnes1992a}, fuelling the growth of central black
holes \citep[e.g.][]{Springel2005} and altering the morphological
mix of the visible Universe
\citep[e.g.][]{Toomre_mergers,Steinmetz2002}. While much of the
literature has focussed on phenomena in the central regions of
merging systems, few studies have, until recently, studied the
impact of the merger process at large distances from the remnant.
Up to a third of the pre-encounter material in the merger
progenitors is tidally ejected during the interaction, into the
tidal tails and bridges that form around the remnant
\citep[e.g.][]{Toomre_mergers,Barnes1992b,Duc1999,Combes1999,Springel1999,Hibbard2005}.
This collisional debris, especially that around gas-rich mergers,
typically hosts star-forming regions, some of which may become
progenitors of self-bound objects with masses typical of dwarf
galaxies
\citep[e.g.][]{Zwicky1956,Schweizer1978,Schombert1990,Mirabel1991,Mirabel1992,Hibbard2005}.
In contrast to normal dwarf galaxies, these `tidal dwarfs' (TDs)
are relatively metal-rich, with metallicities typical of the outer
regions of spiral disks
\citep[e.g.][]{Duc1999,Duc2000,Weilbacher2000,Weilbacher2003},
free of (non-baryonic) dark matter, since their potential wells
are too shallow to capture significant amounts of dark matter
particles \citep[e.g.][but see Gentile et al. 2007, Milgrom
2007]{Bournaud2006,Duc2004} and may contribute a significant
fraction of the nearby dwarf galaxy census
\citep[e.g.][]{Kroupa1997,Hunsberger1996,Okazaki2000,Metz2007}.

Two main mechanisms are postulated for TD formation. Jeans
instabilities within the gas in the tidal tails can lead to
gravitational collapse and the formation of self-bound objects
\citep{Elmegreen1993b}, akin to processes that produce giant
molecular clouds. The Jeans masses are typically high - as the gas
is heated by the merger - enabling the formation of relatively
massive objects, some of which share the properties of local dwarf
galaxies
\citep[e.g.][]{Elmegreen1993b,Struck2005,Bournaud2006,Wetzstein2007,Smith2008}.
Alternatively, a large fraction of the stellar material in the
progenitor disk may be ejected into the outer regions of the tidal
tail, providing a local potential well into which gas condenses
and fuels star formation
\citep[e.g.][]{Barnes1992b,Duc2004,Hancock2009}. In the first
scenario the stellar component is likely to be dominated by young
stars, while in the second a substantial fraction of the stellar
material is expected to be composed of old stars from the disks of
the parent galaxies \citep[see the recent review
by][]{Bournaud2010}. While a rich theoretical and observational
literature has developed on the properties of nearby TDs
\citep[e.g.][]{Wallin1990,Schombert1990,Hibbard1994,Duc2000,Heithausen2000,Braine2001,Hibbard2001,Temporin2003,Mundell2004,Neff2005,Hancock2007,Recchi2007,Bournaud2008,Werk2008,Boquien2009,Sheen2009,Koribalski2009,Boquien2010,Wen2011},
only relatively small samples of TDs have typically been exploited
in any given study. A large statistical study of TDs at low
redshift is clearly desirable.

An impediment to such a study is that a large, statistically
meaningful sample of galaxy mergers in the local Universe has so
far been lacking. This is because, given the small merger fraction
at low redshift \citep[a few percent, see
e.g.][]{Abraham1996,Conselice2003,Lavery2004,de_propris2005,Conselice2008,Darg2010a},
a significant volume of the local Universe must be observed in
order to extract an adequately large sample of merging systems.
While the advent of modern surveys, such as the \emph{Sloan
Digital Sky Survey} \citep[SDSS; ][]{York2000}, has made such data
available, the identification of mergers remains a challenge, both
due to the prodigious size of these datasets and the technical
difficulty in identifying peculiar systems like galaxy mergers.

Automated methods have often been employed to extract mergers from
survey data but most have some limitations. Galaxy `close pairs' -
which are likely progenitors of mergers - can be identified in
spectroscopic surveys
\cite[e.g.][]{Patton2000,LeFevre2000,Nikolic2004,Ellison2008,Rogers2009}.
However, close-pair studies in the SDSS are likely to miss up to
80\% of merging systems because fibre collisions prevent the SDSS
from obtaining spectra for two objects that are within 55
arcseconds of each other in a single visit \citep[see
e.g.][]{Darg2010a}. Quantitative morphological parameters, e.g.
Concentration, Asymmetry, Clumpiness, M$_{20}$ and the Gini
coefficient, have been extensively employed (often through the use
of neural networks) to classify galaxy morphologies in large
surveys
\citep[e.g.][]{Abraham1996,Conselice2000,Abraham2003,Conselice2003,Ball2004,Lotz2004,Ferreras2005,Lahav1995,Lisker2008,Andrae2011}.
However, it is difficult to define a parameter space that is
\emph{unique} to mergers and the results of such quantitative
methods are typically checked and calibrated against visual
inspection \citep[e.g.][]{Abraham1996,Ferreras2009,Jogee2009},
which is arguably the most reliable method of morphological
classification. The utility of visual inspection becomes
particularly important for identifying peculiar systems, such as
ongoing mergers and post-mergers
\citep[e.g.][]{Cassata2005,Kaviraj2010b}. However, since it is
prohibitively time-consuming for large datasets, visual inspection
of the SDSS has, until the advent of the \emph{Galaxy Zoo} (GZ)
project, been limited to very small fractions (a few percent or
less) of the full spectroscopic galaxy sample in this survey
\citep{Fukugita2007,Schawinski2007a,Nair2010}.

\begin{figure*}
\begin{minipage}{172mm}
\begin{center}
$\begin{array}{ccc}

\includegraphics[width=2.2in]{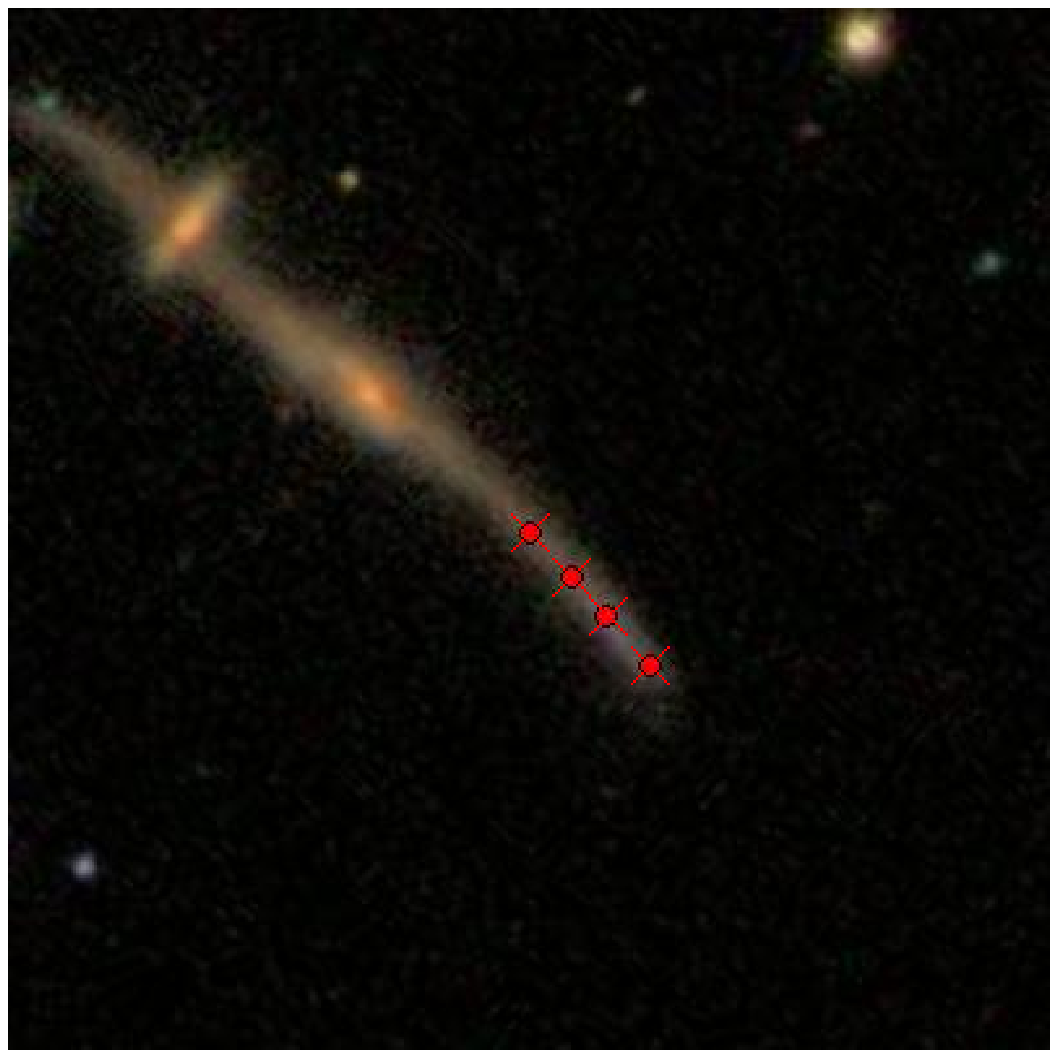} & \includegraphics[width=2.2in]{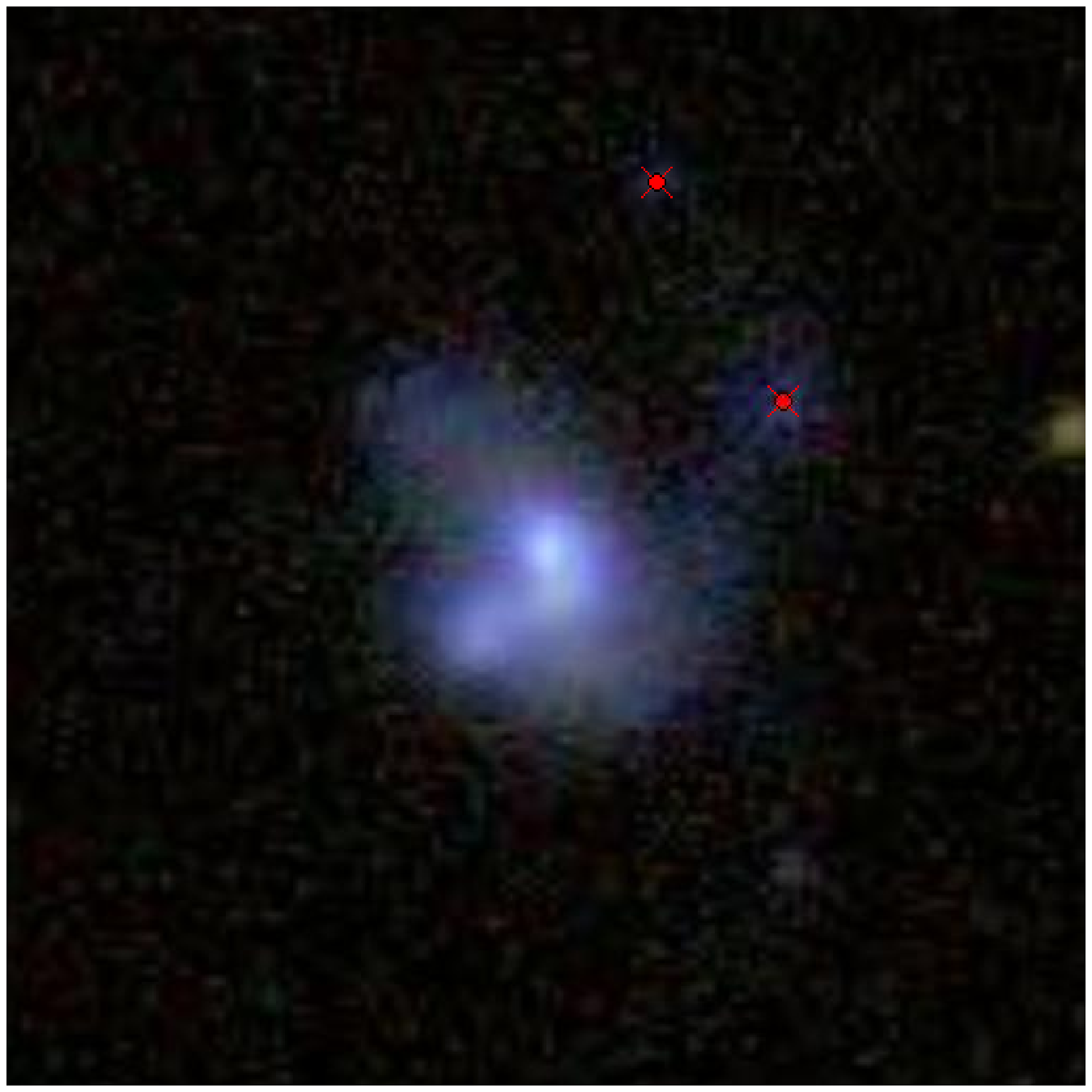} &
\includegraphics[width=2.2in]{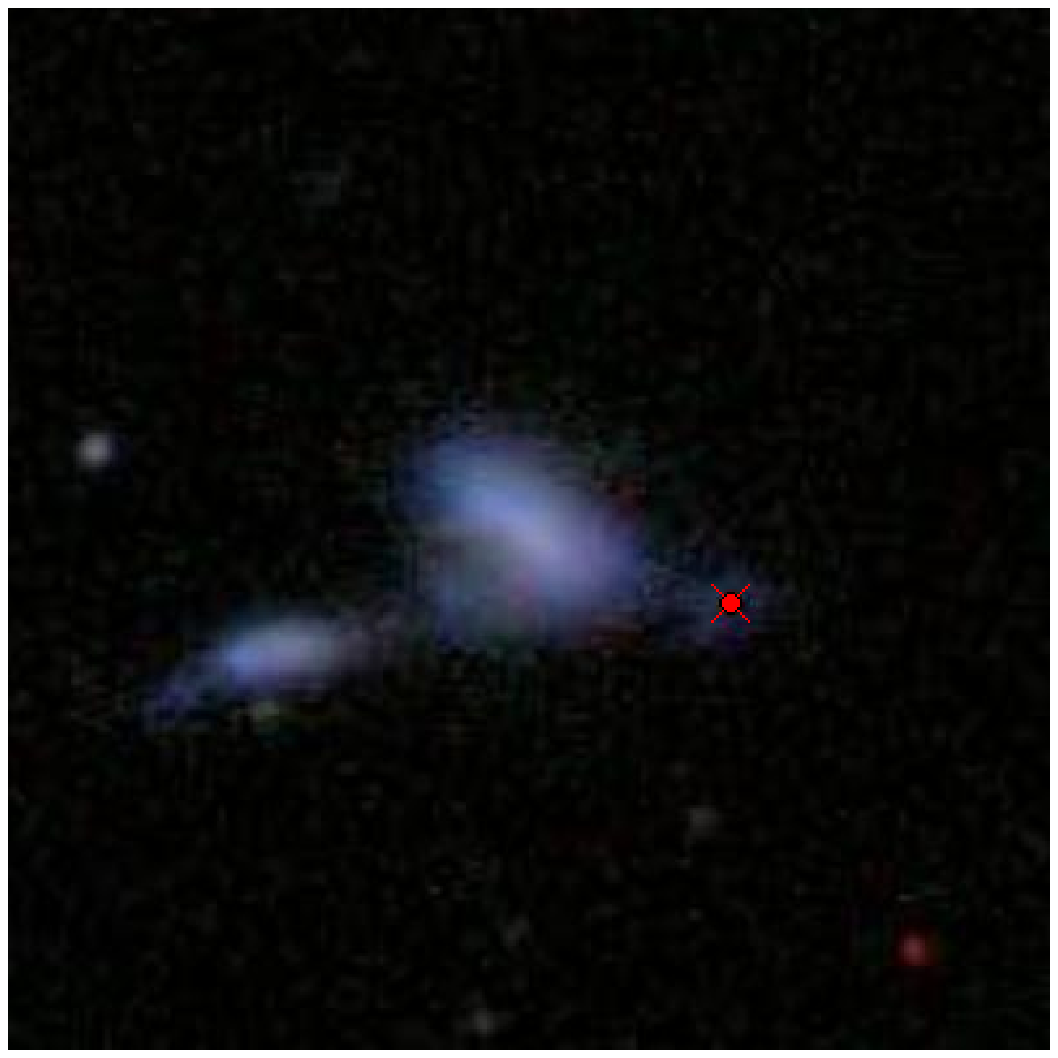}\\

\includegraphics[width=2.2in]{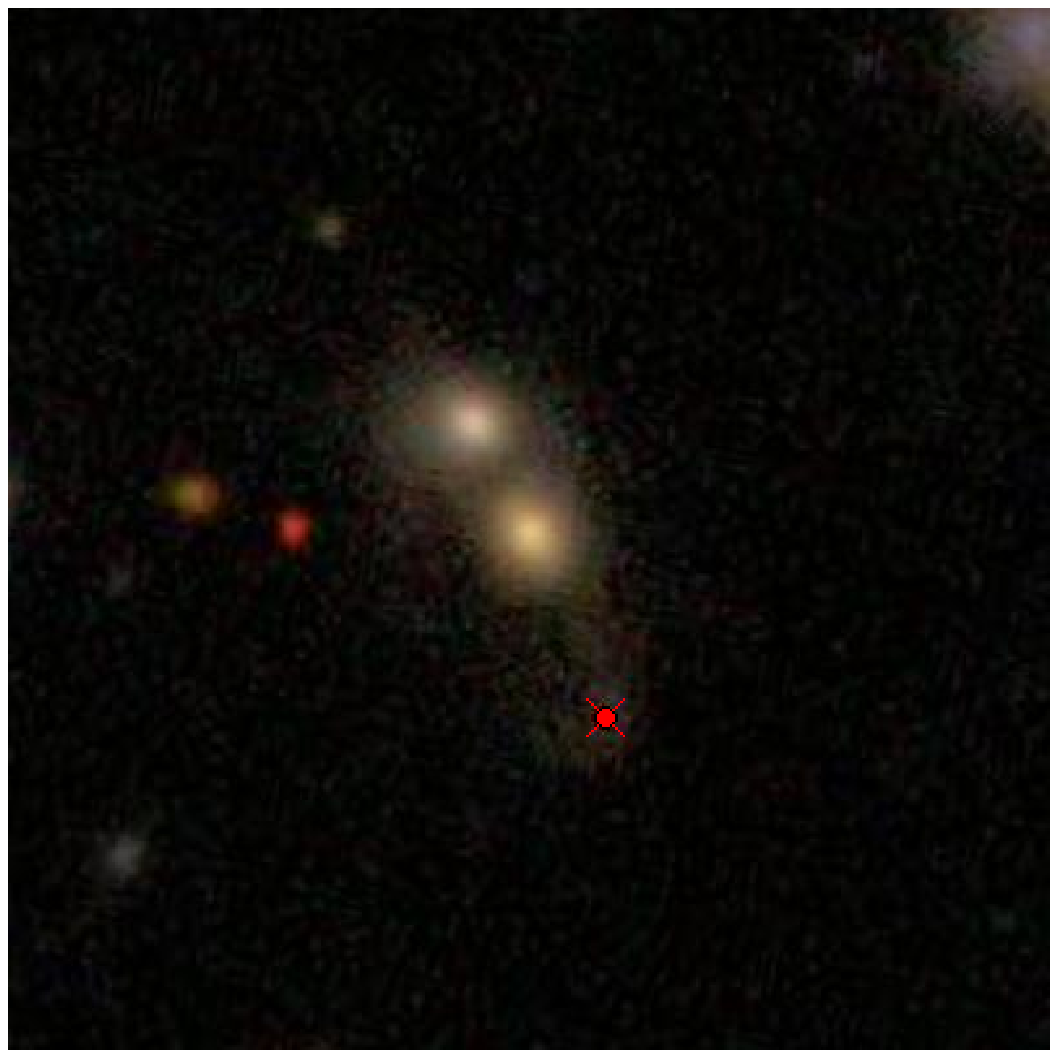} & \includegraphics[width=2.2in]{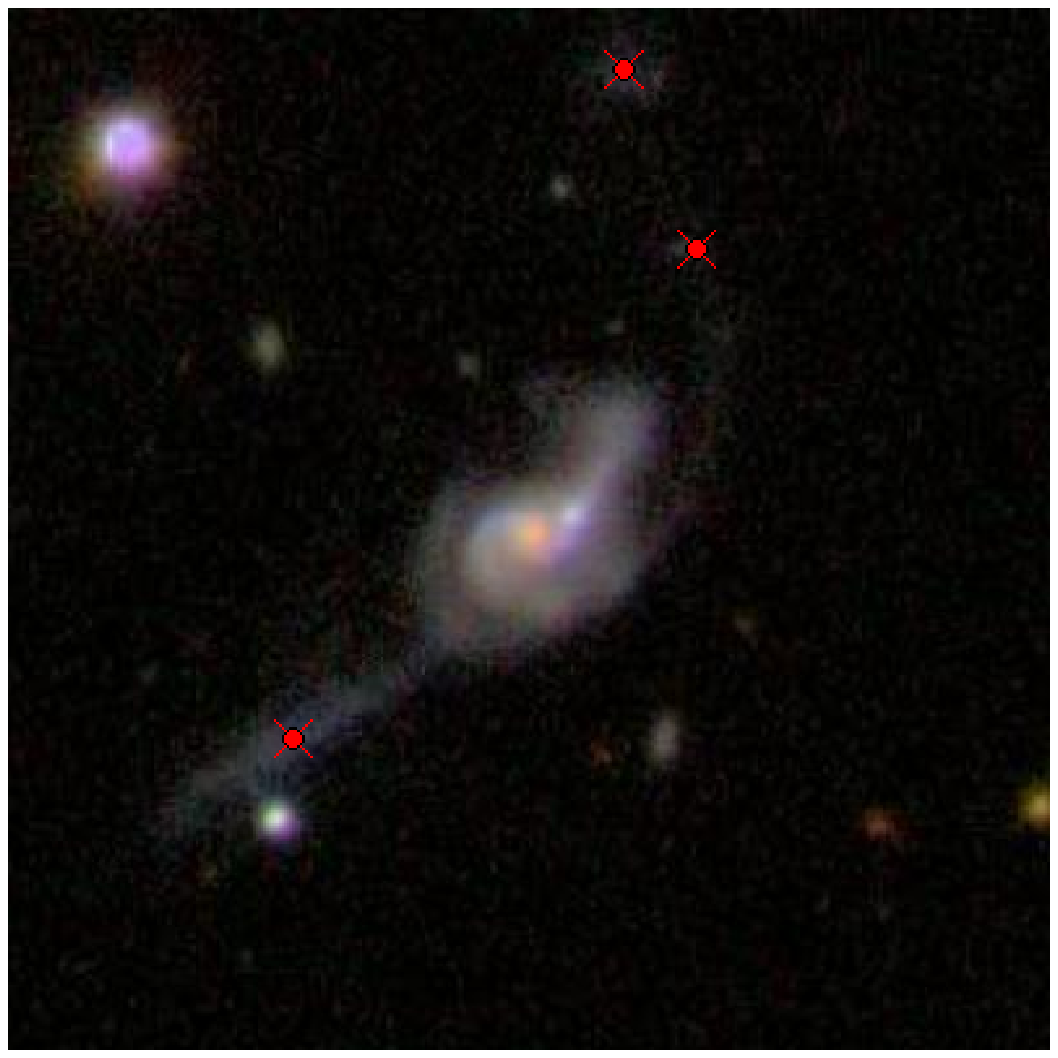} &
\includegraphics[width=2.2in]{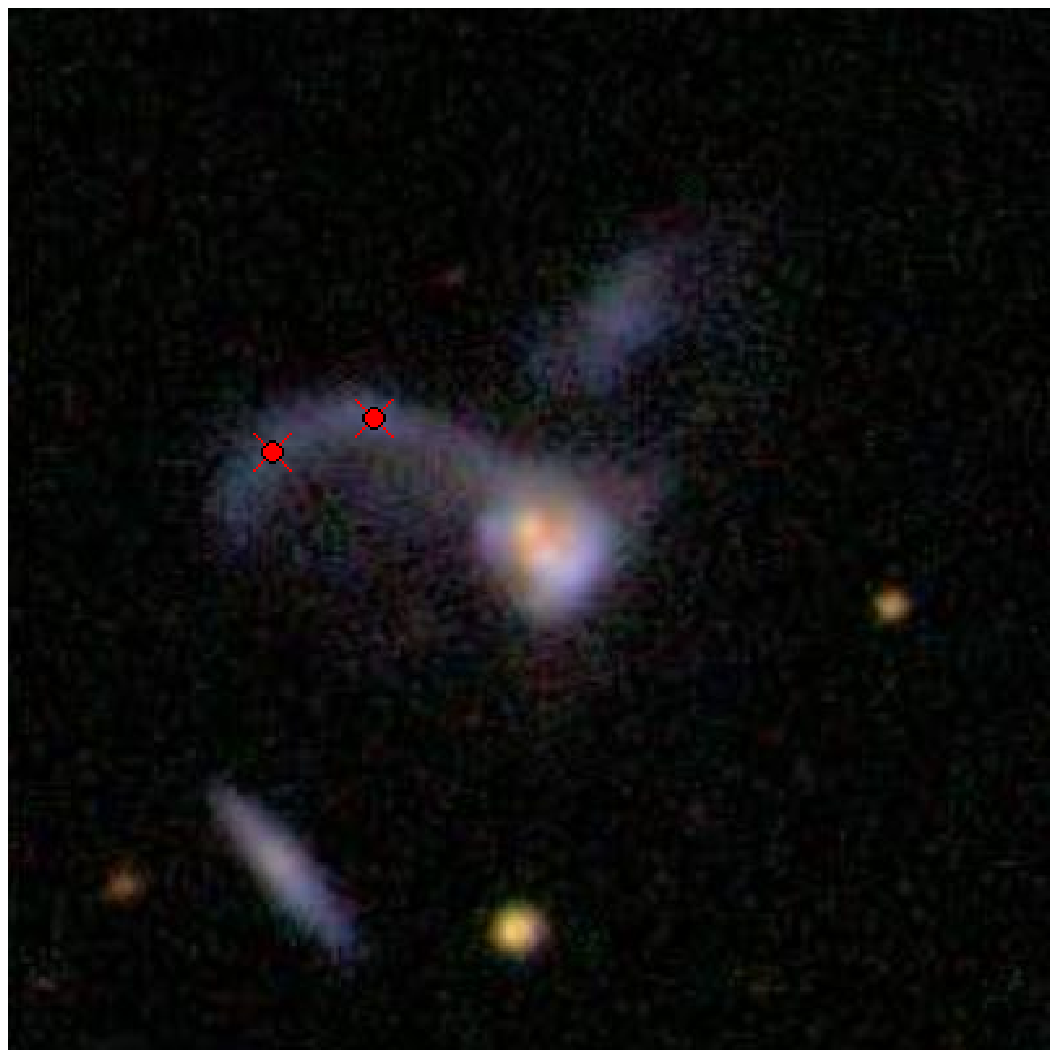}\\

\end{array}$
\caption{Examples of tidal dwarf candidates in the Galaxy Zoo
mergers sample. Tidal dwarfs are selected as separate photometric
objects, identified by the SDSS pipeline, that are clearly
associated with the tidal debris around the merger in question.
The positions of these objects are marked by the red crosses. This
figure is available in colour in the online version of the
journal.} \label{fig:td_examples}
\end{center}
\end{minipage}
\end{figure*}

GZ is a citizen-science project which has used 250,000+ volunteers
from the general public to morphologically classify the entire
SDSS spectroscopic sample ($\sim$1 million galaxies) through
visual inspection of their optical images
\citep{Lintott2008,Lintott2011}. In particular, it offers an
unprecedented route to extract a statistically meaningful sample
of galaxy mergers in the local Universe. \citet{Darg2010a} have
used the GZ database, based on the SDSS Data Release 6
\citep{SDSSDR6}, to construct a robust sample of
{\color{black}3373} mergers with redshifts less than 0.1, typical
mass ratios between 1:1 and 1:10 and a wide variety of
separations, ranging from systems that are `on approach' to ones
that are almost fully coalesced. We refer readers to
\citet{Darg2010a,Darg2010b} for details of how this sample was
constructed and the general properties of the galaxy mergers
therein.

In this paper we exploit the Darg et al. merger sample to study
the local TD population. The aim is to complement existing TD
studies, which are typically based on relatively small samples of
galaxy mergers, by offering a statistical view of the properties
of the TD population in our local neighbourhood. The plan for this
paper is as follows. In Section 2 we describe the compilation of a
TD sample from the Darg et al. study. In Section 3 we study the
general properties of TD-producing systems, cataloguing the
morphologies, mass ratios and local environments of the parent
mergers that produce TDs. In Section 4 we compare the properties
of TDs (e.g. masses and colours) to their parents and explore the
impact of Active Galactic Nuclei (AGN) in the parent galaxies on
the formation of TDs in the tidal tails. In Section 5, we explore
the star formation histories of the TDs. In particular, we
investigate the ages and mass fractions of young stars in
individual TDs, in order to compare the fraction of new stellar
mass that is born in situ with that composed of old stars drawn
from the parent disks. Finally, in Section 6, we explore whether
the TD population could make a significant contribution to the
dwarf galaxy census in nearby clusters. We summarise our findings
in Section 7.

%...................................................................................................

\section{Compiling a sample of tidal dwarfs}
TDs are identified through visual inspection of the co-added
$g,r,i$ SDSS images of each merger. Separate photometric objects,
extracted by the SDSS pipeline, that are clearly associated with
the tidal debris in each merger are selected as TDs. In $\sim$20\%
of the cases there are multiple photometric objects associated
with the same TD, with one object typically containing more than
90\% of the flux. In such cases we sum the fluxes of all the
photometric objects to estimate the flux of the TD in question. We
note that if we simply used the brightest photometric object in
each of these cases the general conclusions of our study would
remain unaffected. This procedure yields 405 TDs. For each TD we
also record an approximate position in relation to their parent
galaxy - at the tidal-tail tips, within the tail or at the base of
the tidal tail.

Figure \ref{fig:td_examples} presents examples of TDs in our
study. The position of the individual photometric objects,
identified by the SDSS pipeline, that are selected as TDs are
indicated on the images by red crosses. It should be noted that
the identification of these objects relies on their association
with the tidal debris in the parent mergers. However, since the
objects are still `attached' to their parents (which allows us to
identify them as potential TDs in the first place), we cannot be
certain whether they will evolve into independent self-bound
objects and eventually contribute to the dwarf galaxy population.
Hence the objects identified here are, strictly speaking, \emph{TD
candidates}.

The number of TDs per merger does not evolve across our redshift
range ($0.01<z<0.1$), which suggests that the TD population
identified at the lower end of our redshift range is similar to
that identified at the upper end. In other words, we expect the TD
population to be relatively homogeneous across the redshift range
of this study. Since they are, by definition, associated with
their parent merger, we calculate TD redshifts from the
spectroscopic redshift information available for the parent
galaxies. Due to fibre collisions the SDSS does not measure
spectra for two objects that are within 55 arcseconds of each
other in a single visit. Hence, in the overwhelming majority
(80\%) of systems in the Darg et al. sample, only one merger
progenitor has a spectroscopic redshift. In these cases we take
this as the redshift of all TDs associated with the merger. In
cases where both progenitors have measured spectroscopic
redshifts, we take their average as the redshift of the TDs
belonging to that system. Since the two redshifts are very
similar, this averaging procedure does not affect our results. The
median redshift of the TDs studied in this paper is $z \sim 0.05$.

\begin{figure}
$\begin{array}{c}
\includegraphics[width=3.5in]{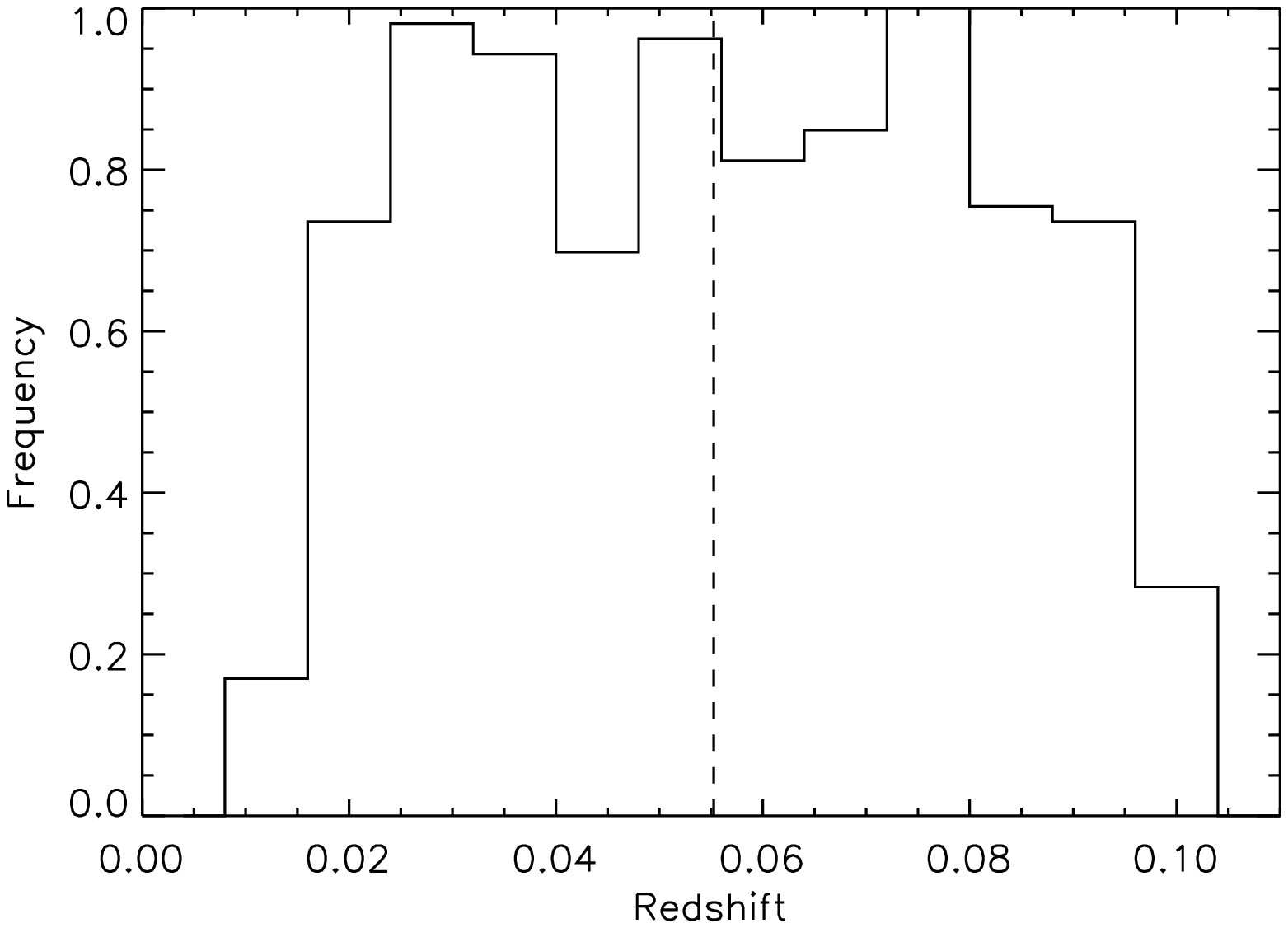}\\
\includegraphics[width=3.5in]{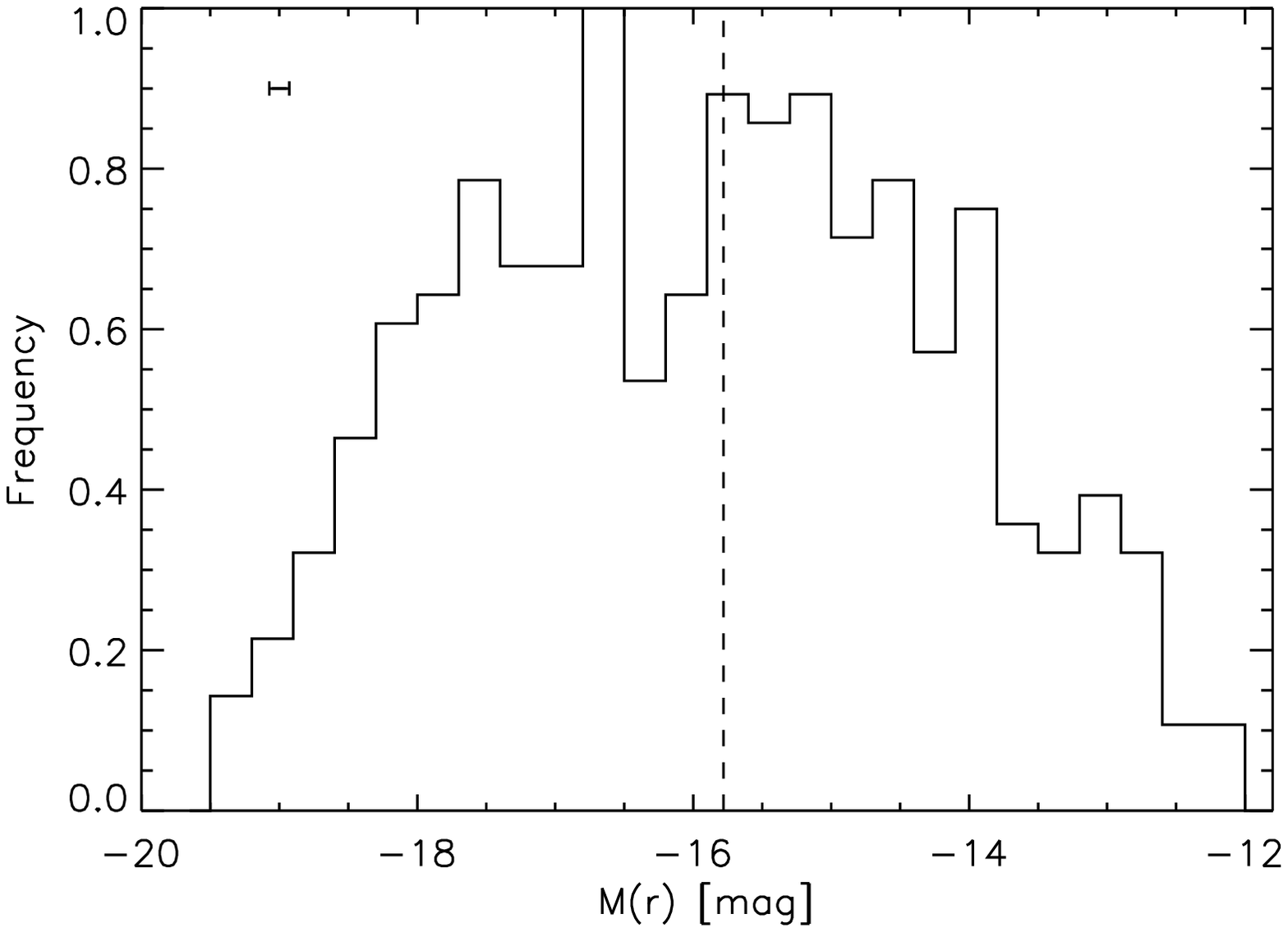}\\
\includegraphics[width=3.5in]{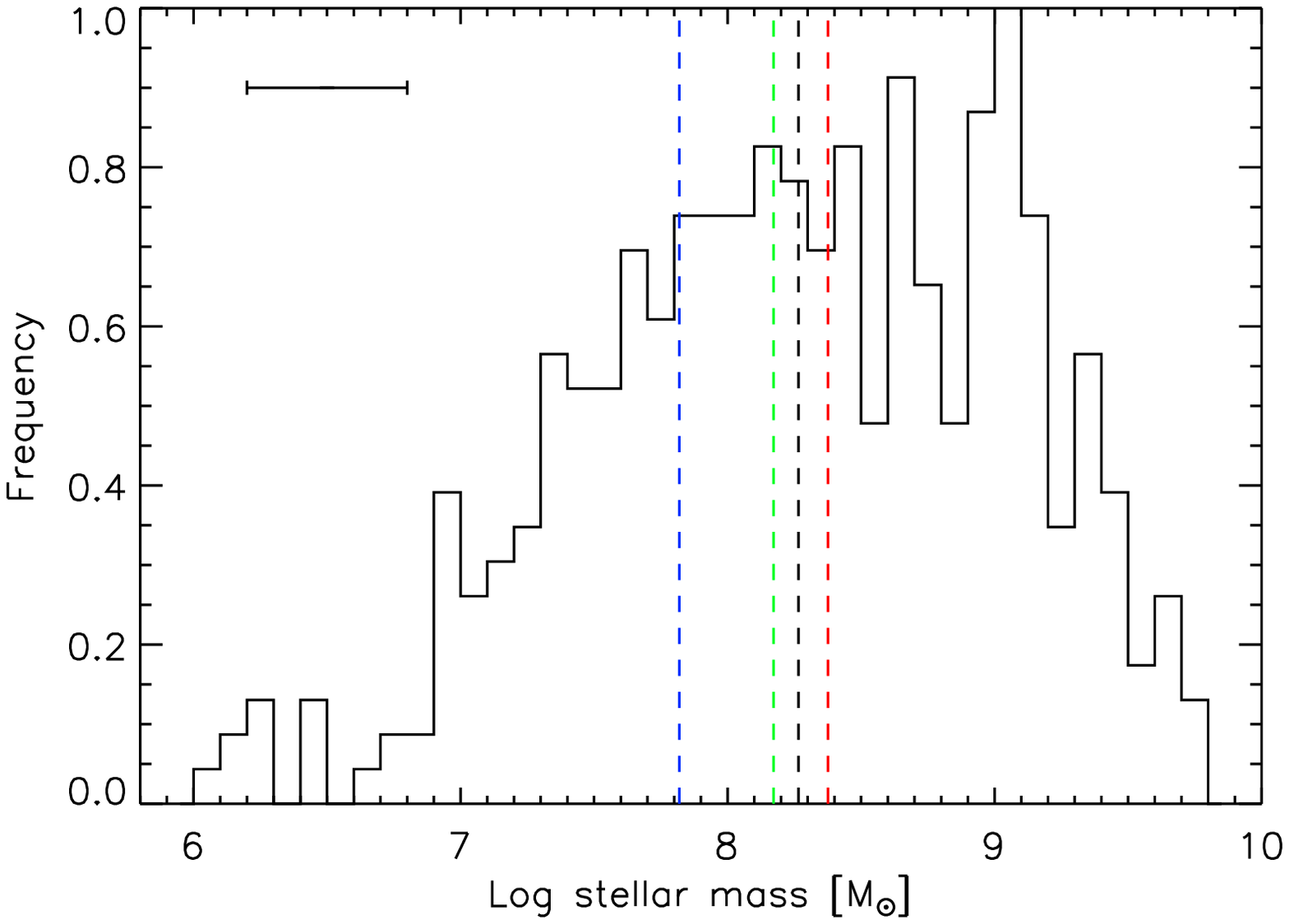}
\end{array}$
\caption{Distributions of redshift {\color{black}(top)}, $r$-band
absolute magnitude {\color{black}(middle)} and stellar mass
{\color{black}(bottom)} for the TD sample in this study. The
stellar masses are calculated using the calibrations of
\citet{Bell2003} and have errors of up to $\sim$0.3 dex. Median
values of individual distributions are indicated using the
vertical dashed lines. In the bottom panel (stellar masses) three
additional vertical lines are shown which indicate median values
for TDs at the tips of tidal tails (red), along the tails (green)
and at the base of tails (blue). Errors in spectroscopic redshifts
are negligible ($\sim10^{-4}$). Magnitude errors are taken from
the SDSS DR6 database. This figure is available in colour in the
online version of the journal.} \label{fig:general_props1}
\end{figure}

The redshifts are used to calculate absolute magnitudes for each
TD, from the apparent magnitudes measured by the SDSS pipeline.
K-corrections are computed using the latest version of the public
\texttt{KCORRECT} code \citep{Blanton2003a,Blanton2007}. The
absolute magnitudes are used to estimate stellar masses, using the
calibrations of \cite{Bell2003}. The error on these masses can be
up to 0.3 dex. Figure \ref{fig:general_props1} presents
distributions of the basic properties (redshift, absolute $r$-band
magnitude and stellar mass) of the TD sample in this paper. Median
values are indicated using the dashed vertical lines. Note that,
in the bottom panel (stellar mass), three additional vertical
lines are shown, which indicate median values for TDs at the tips
of tidal tails (red), within the tails (green) and at the base of
tails (blue). We return to these TD subsets in Section 3.3 below.

\begin{comment}
It is instructive to explore the completeness of the TD sample
studied here. The identification of TDs clearly depends on the
visibility of the tidal tails. If the visibility of the tidal
debris evolves significantly across our redshift range then it is
possible that our ability to detect decreases at the the higher
end of this redshift range.

Since the Darg et al. sample is expected to be unbiased in terms
of merger type across this redshift range, it appears reasonable
to assume that the TDs studied here are representative of the TD
population in the local Universe.

Since the merger fraction is very small and there are a few TDs
per merger it seems unlikely that the bulk of the dwarf galaxy
population are tidal in origin.
\end{comment}

%...................................................................................................

\section{Properties of the parent mergers}
\subsection{Parent morphologies and colours}
We begin by cataloguing the properties of TD-producing mergers and
compare them to the general merger population. 95\% of binary
mergers that produce TDs involve two spiral progenitors, while
most remaining systems have at least one spiral progenitor. The
fraction of TD-producing mergers where both parent galaxies have
early-type morphology is less than 2\% (at least in the sample
studied here), strongly suggesting that TDs are unlikely to form
in {\color{black}such} mergers. It is instructive to check whether
the significant lack of TDs in {\color{black}early-type -
early-type (E-E)} mergers is a real effect or whether they are not
identified (partly) because the tidal tails in these mergers are
too faint to be clearly detected in the standard SDSS imaging. By
virtue of being a large-area survey, the standard SDSS imaging is
relatively shallow, with only $\sim$54 second exposures in every
filter (the $r$-band detection limit is $\sim$22 mag). To probe
this issue further we explore the images of {\color{black}E-E}
mergers in this sample that lie in the SDSS Stripe 82
$(-50^{\circ} < \alpha < 59^{\circ}, -1.25^{\circ} < \delta <
1.25^{\circ})$ that offers 2 mag deeper imaging than the standard
SDSS survey. The Stripe 82 has been imaged multiple times as part
of the SDSS Supernova Survey \citep{Frieman2008} and achieves
limiting magnitudes of $\sim24$ mag in $r$-band, sometimes
{\color{black}revealing faint tidal debris in mergers that may be
invisible in the standard imaging} \citep{Kaviraj2010a}. Since it
has an area of 270 deg$^2$, compared to the 9583 deg$^2$ area of
the DR6 from which the Darg et al. sample is constructed, only 9
{\color{black}E-E} mergers\footnote{{\color{black}The total number
of mergers in the Darg et al. sample is 3373, with an E-E fraction
of around 12\%. The area of the Stripe 82 is 3\% of the DR6 (270
deg$^2$/9583 deg$^2$). Thus we expect ($270 \times 3373 \times
0.12/9583 \sim 11$) E-E mergers in the Stripe 82. The actual
number is 9 (consistent within counting errors).}} in the Darg et
al. study lie in this region of the sky. However, visual
inspection of these images do not reveal any TDs not identified in
the standard images, leaving our conclusions above unchanged.

Taking $(g-r) \sim 0.7$ as the transition between the blue cloud
and the red sequence \citep[see
e.g.][]{Strateva2001,Blanton2003b}, we find that in 85\% of the
parent mergers both progenitors are blue. In 12\% at least one
progenitor is blue, while in the remaining 3\% of parent systems
both progenitors are on the red sequence. Not unexpectedly TD
formation becomes significantly more likely when both merger
progenitors are gas-rich (and therefore in the blue cloud).

%...................................................................................................

\subsection{Parent mass ratios}
\begin{figure}
\includegraphics[width=3.5in]{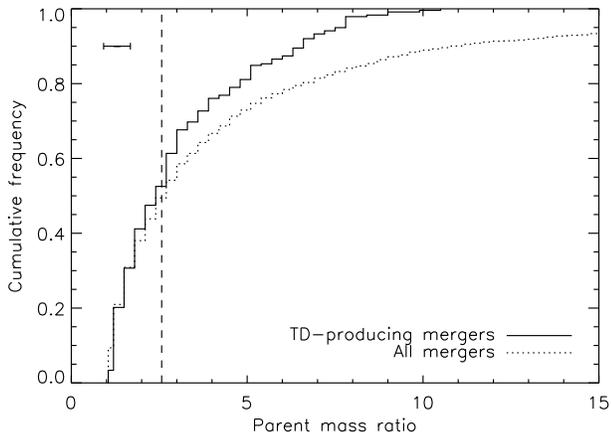}
\caption{Parent mass ratios of TD-producing mergers (solid line)
compared to the general merger population (dotted line). 95\% of
TD-producing mergers have mass ratios greater than $\sim$1:7. The
median mass ratio of TD-producing mergers is $\sim$1:2.5 (shown
using the vertical dashed line).} \label{fig:parent_massratios}
\end{figure}

Figure \ref{fig:parent_massratios} indicates that 95\% of TDs are
produced by parent systems whose constituent galaxies have mass
ratios greater than $\sim$1:7. The median parent mass ratio is
$\sim$1:2.5. TDs are not produced by systems where the mass ratio
exceeds $\sim$1:11. TD formation, in other words, appears much
less likely in minor mergers (typically defined as systems with
parent mass ratios less than 1:4). These observational results
support recent theoretical work, which suggests that favourable
conditions for TD formation require gas-rich mergers with mass
ratios greater than 1:8 \citep{Bournaud2006}.

%...................................................................................................

\subsection{Separation of tidal dwarfs from parent galaxies}
\begin{figure*}
\begin{minipage}{172mm}
$\begin{array}{cc}
\includegraphics[width=3.5in]{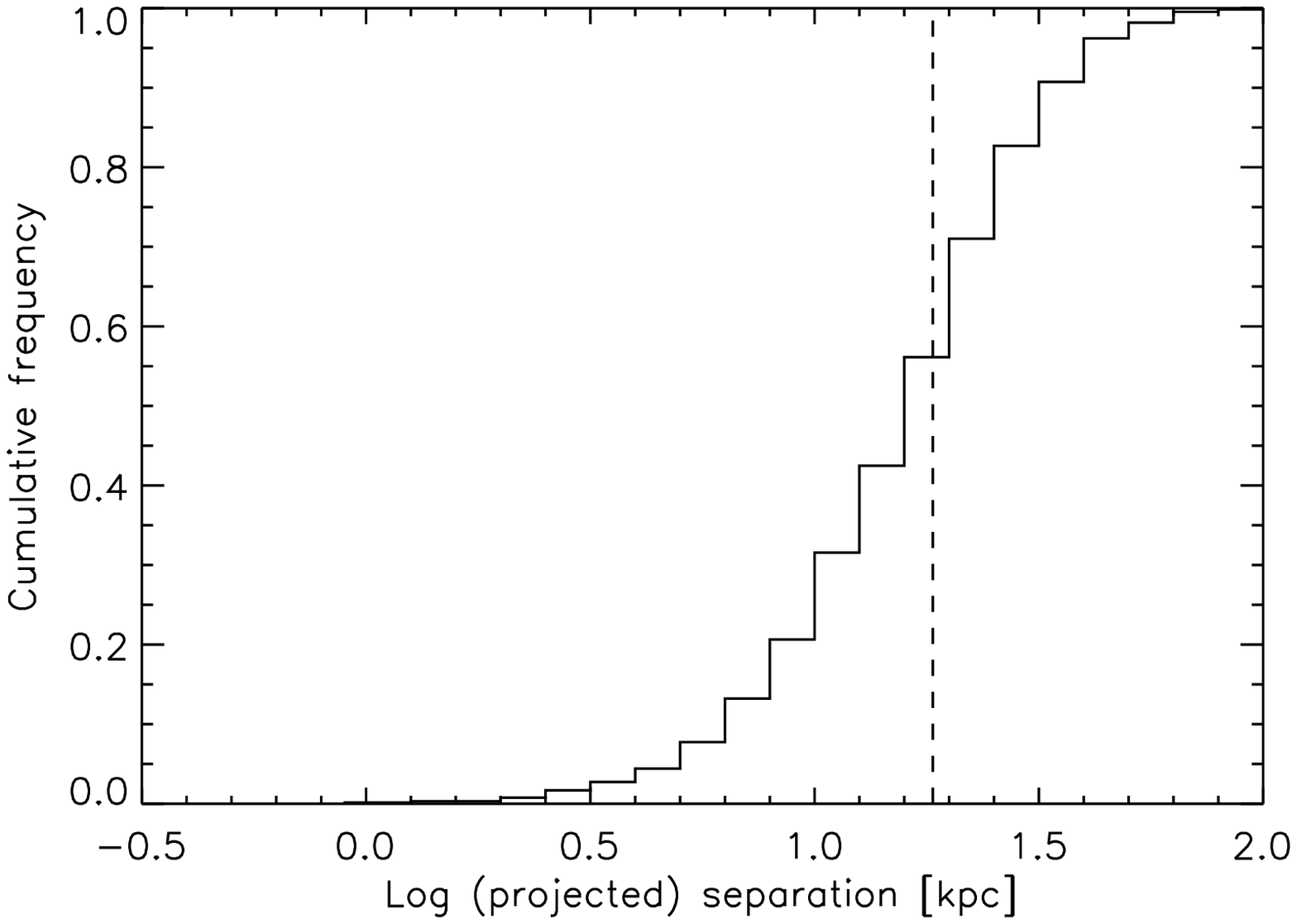} & \includegraphics[width=3.5in]{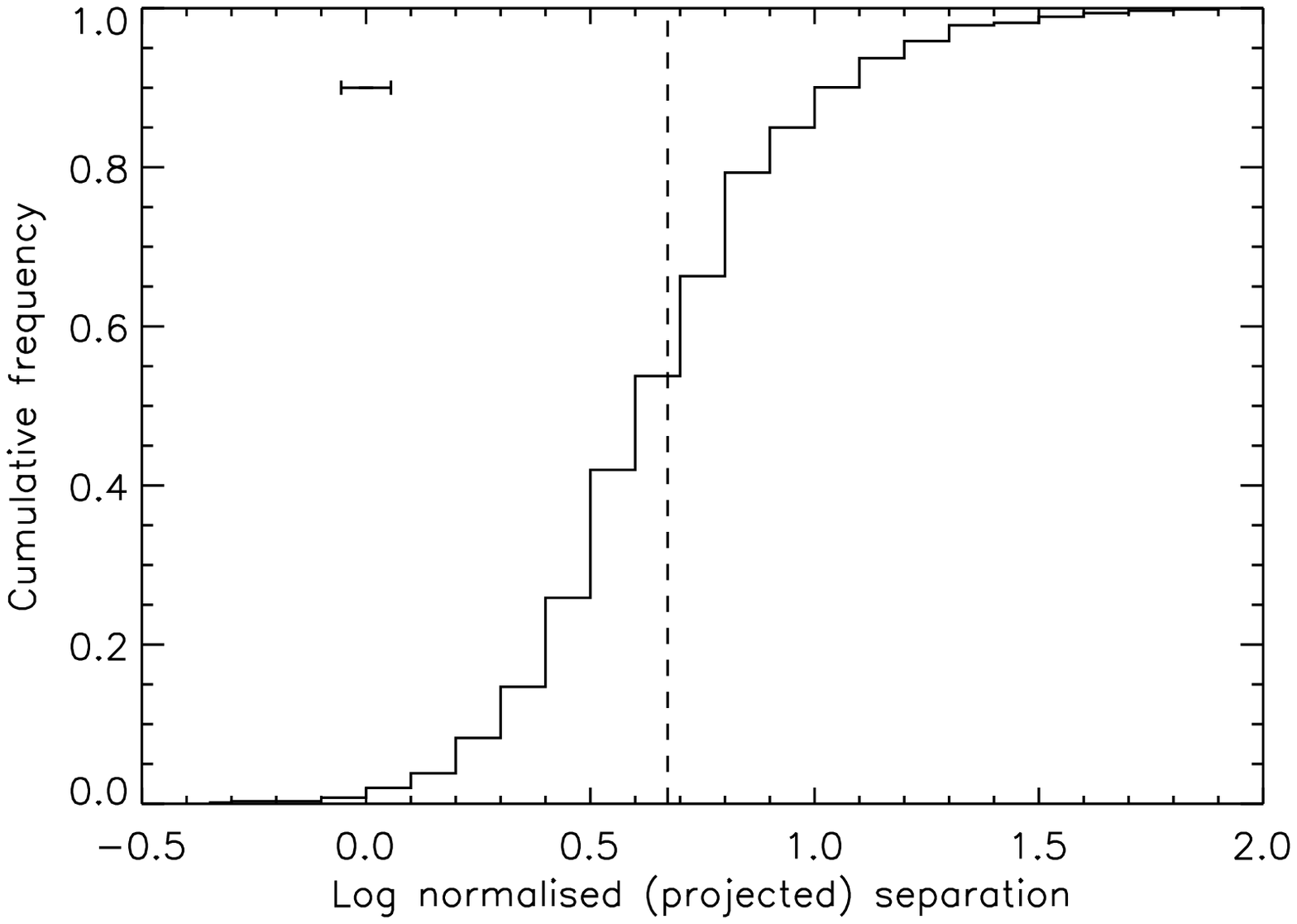}
\end{array}$
\caption{\textbf{LEFT:} Physical separation of TDs from their
parent galaxies. \textbf{RIGHT:} Physical separation normalised by
the half-light radius ($R_{1/2}$) of the parents. Median values
are indicated using dashed vertical lines. The median separation
is $\sim$17 kpc and the median normalised separation is $\sim 4.5
R_{1/2}$. 95\% of TDs are within $\sim 15 R_{1/2}$ of their parent
galaxies. We assume that the uncertainties in RA and DEC values
are negligible (hence no error bar is shown on the separations).
The error in the normalised separation (right-hand panel) is
driven by the error in the half-light radii of the parent
galaxies.} \label{fig:td_separations}
\end{minipage}
\end{figure*}

In Figure \ref{fig:td_separations} we show both the physical
separation of TDs from their parents (left panel) and the
separation normalised by the half-light radii ($R_{1/2}$) of the
parent galaxies. 95\% of TDs are within $\sim$20 kpc of their
parent galaxies, corresponding to $\sim 15 R_{1/2}$ (the median
normalised separation is $\sim$17 kpc or $\sim 5 R_{1/2}$),
generally consistent with the theoretical simulations of
\citet{Bournaud2006}.

As indicated in the bottom panel of Figure
\ref{fig:general_props1} above, TDs that lie further along the
tidal tail appear to be more massive. {\color{black}The offset in
the median masses (indicated by the dashed lines) of TDs born at
the tips of the tidal tails compared to those born at the base of
the tails} is $\sim$0.6 dex (around a factor of 4). This mass
offset is expected because the most massive objects are likely to
form at the tail tips where the tidal debris accumulates
\citep{Elmegreen1993b,Bournaud2004,Duc2004}.

%Show the separation vs mass plot. Mention that there is a weak trend of the TD mass scaling with the
%distance from the parent. This is consistent with that is found in Hunsberger et al. but the trend is
%much weaker in this study. Note also that Weilbacher et al. did not find any trend so this suggests
%that the situation varies quite a lot depending on the parent systems in question.

%...................................................................................................

\subsection{Local environment}
\begin{figure}
\includegraphics[width=3.5in]{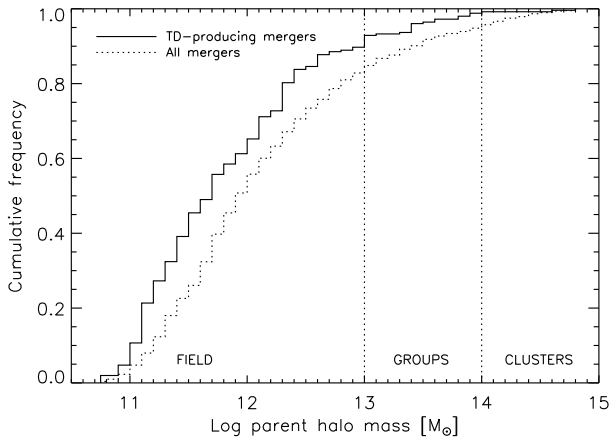}
\caption{Local environments of TD-producing mergers (solid line)
compared to the general merger populations (dotted line).
TD-producing mergers typically inhabit field environments. Note
that the \citet{Yang2007} catalogue, from which the environment
measures are extracted, does not provide any error information on
the host dark-matter halo masses.} \label{fig:td_environments}
\end{figure}

We explore the local environment of TD-producing mergers by
cross-matching with the SDSS environment catalogue of
\citet{Yang2007,Yang2008}, who use a halo-based group finder to
separate the SDSS into 300,000+ structures, spanning rich clusters
to the field. The catalogue provides estimates for the masses of
the host dark matter haloes of individual SDSS galaxies, which are
related to the traditional classifications of environment
(field/group/cluster). Haloes with masses greater than
$10^{14}$M$_{\odot}$ represent clusters, while those with masses
in the range $10^{13}$M$_{\odot}$ to $10^{14}$M$_{\odot}$
represent groups. Smaller DM haloes represent the field. Figure
\ref{fig:td_environments} indicates that TD-producing mergers
favour lower-density environments than the general merger
population. $\sim$90\% of TD-producing mergers reside in the
field, with the remaining systems inhabiting groups. Almost none
of the systems reside in clusters. This result is consistent with
the fact that the availability of cold gas is a strong function of
local environment. A cluster environment, in particular, is
expected to be cold-gas-poor \citep[e.g.][]{Solanes2001} and
therefore hostile to TD formation.

%...................................................................................................

\section{Tidal dwarf vs. parent properties: masses, colours and the impact of AGN}
\begin{figure}
\includegraphics[width=3.5in]{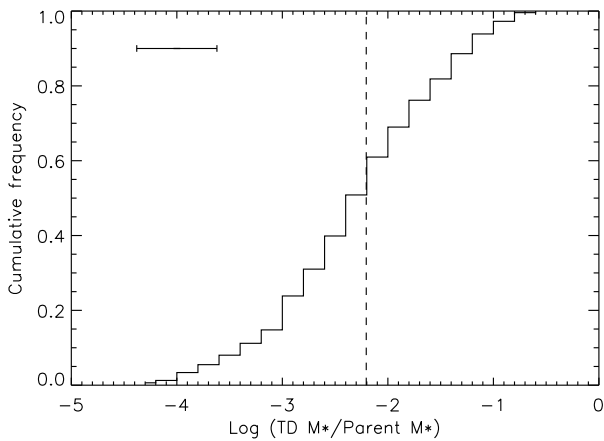}
\caption{Ratio of TD stellar mass to parent stellar mass. 95\% of
TDs have stellar masses that are less than $\sim$10\% of the
stellar masses of their parent galaxies. The median stellar mass
ratio is $\sim$0.6\%, indicated by the dashed vertical line.}
\label{fig:tdtoparent_massratios}
\end{figure}

We proceed by comparing how TD properties compare to those of
their parent mergers. Figure \ref{fig:tdtoparent_massratios}
indicates that the stellar masses of 95\% of TDs are less than
10\% of the stellar mass of their parent mergers. The median
TD-to-parent stellar mass ratio is around 0.6\% (shown using the
dashed line in Figure \ref{fig:tdtoparent_massratios}). Note that,
since the masses are calculated from the photometric data, they
correspond only to the stellar component of the galaxy. While the
dynamical (total) masses of the TDs are expected to be similar to
their stellar masses (since they do not contain significant
amounts of dark matter), this is not the case for the parent
spiral galaxies, which may contain {\color{black}3-5 times as much
dark as luminous matter
\citep[e.g.][]{vanalbada1986,Ashman1992,Salucci2000,Noordermeer2007,Salucci2009}
inside $\sim$10 disk scalelengths (typically 20-40 kpc)}. The
\emph{total} TD-to-parent mass ratios are therefore likely to be
several factors smaller than the stellar values derived here.

We now compare the TD colours to those of their parent galaxies.
Recent studies have suggested that the presence of an AGN in a
galaxy can affect the colours of objects in their immediate
vicinity \citep{Shabala2011}, plausibly due to interaction between
AGN-driven outflows and the gas reservoirs of these nearby
galaxies. It is conceivable, therefore, that TD formation might
also be affected by outflows due to nuclear activity in their
parents. This may either suppress star formation by removing gas
from the star-forming regions, as is typically envisaged in
negative feedback scenarios
\citep[e.g.][]{Silk1998,Croton2006,Tortora2009,Kaviraj2010c} or
alternatively enhance star formation through positive feedback by
compressing cloud complexes
\citep[e.g.][]{vanbreugel1985,Mould2000,Silk2005} in the tidal
tails. Thus, we also wish to explore the colour difference between
TDs and parents as a function of AGN activity in the parent
galaxies.

Parent galaxies that host AGN are identified using a standard
optical emission-line ratio analysis
\citep[e.g.][]{Baldwin1981,Veilleux1987,Kauffmann2003,Kewley2006},
which classifies objects as `star-forming', `composite' (which
have signatures of both AGN and star formation), `Seyfert',
`LINER' or `quiescent'. The analysis is performed using the public
\texttt{GANDALF} code \citep{sauron5}\footnote{GANDALF
simultaneously fits the emission and absorption lines and is
designed to separate the relative contribution of the stellar
continuum and of nebular emission in the spectra of nearby
galaxies, while measuring the gas emission and kinematics. See
http://star-www.herts.ac.uk/$\sim$sarzi/PaperV\_nutshell/PaperV\_nutshell.html
for more details.}. We assume that galaxies classified as
composite, Seyfert or LINER host AGN.

\begin{figure}
\includegraphics[width=3.5in]{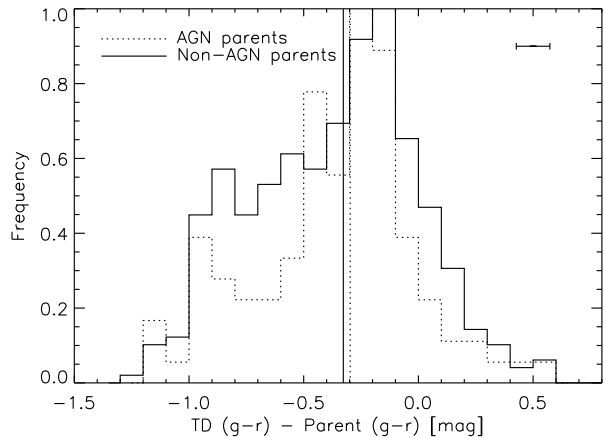}
\caption{TD $(g-r)$ colour vs. parent $(g-r)$ colour. We show
parents with and without AGN using the vertical lines. AGN are
identified using an optical emission-line-ratio analysis (see text
in Section 4 for details). Median values are indicated using
vertical lines.} \label{fig:tdtoparent_colours}
\end{figure}

Figure \ref{fig:tdtoparent_colours} shows that TDs are typically
bluer than their parents. The median $(g-r)$ colour offset between
TDs and parents is $\sim$0.3 mag, similar to results of previous
optical studies \citep[e.g.][]{Duc1999} based on smaller TD
samples. We also find that parents hosting AGN do not show a
larger colour difference (within the errors) from their TDs than
those without AGN. It seems unlikely, therefore, that feedback
from AGN (either positive or negative) plays a role in TD
evolution.

%...................................................................................................

\section{Tidal dwarf star formation histories}
We investigate the star formation histories (SFHs) of TDs, in
particular the relative fraction of stellar mass that is composed
of new stars compared to the fraction that is constituted by old
stars from the parent disks. We estimate the SFH of each TD by
comparing colours constructed from the SDSS $(u,g,r,i,z)$
magnitudes to a library of synthetic photometry that is based on
model SFHs designed to approximate the stellar content of each TD.

Each model SFH is constructed using two instantaneous starbursts.
The first burst, which characterises the old, underlying stellar
population in the parent disks, is assumed to have an age of 7
Gyr, which represents an \emph{average} age for the old disk
stars. Recent studies that have decoupled the recent star
formation from the old, underlying populations in star-forming
spirals suggest average ages for the old stars around this value
\citep[see][]{Kaviraj2009}. We have checked that our conclusions
remain unaffected if we change the age of the old stars to 10
Gyrs.

The second burst, which represents the young stellar content of
the TDs, is allowed to vary in (i) age between 1 Myr and 7 Gyr and
(ii) mass fraction between 0 and 1. We also include a range of
values for the internal dust extinction, parametrised in terms of
$E_{B-V}$ from 0 to 1. The dust extinction is applied using the
empirical law of \citet{Calzetti2000} to the SFH as a whole. We
assume that the model SFHs have a metallicity of 0.3$Z_{\odot}$,
which is typical of the outer regions of spiral disks
\citep[e.g.][]{Duc1998,Weilbacher2000}. The free parameters are,
therefore, the age ($t_2$) and mass fraction ($f_2$) of the second
burst and the dust content ($E_{B-V}$) of the TD.
{\color{black}The model SFH library contains 1.5 million
individual models.} To build a library of synthetic photometry,
each combination of free parameters is combined with the stellar
models of \citet{Yi2003} and convolved with the correct SDSS
filtercurves. Since our TD sample spans a range in redshift
($0.01<z<0.1$), equivalent libraries are constructed at redshift
intervals $\delta z=0.005$.

{\color{black}For each TD in our sample, the free parameters are
estimated by choosing the model library that is closest to it in
redshift and comparing its ($u,g,r,i,z$) colours to every model in
the synthetic library. In a Bayesian framework \citep[see
e.g.][]{Sivia2007}, for a vector $\textbf{X}$ denoting parameters
in the model and a vector $\textbf{D}$ denoting the measured
observables (in this case the colours),

\begin{equation}
\textnormal{prob}(\textbf{X}|\textbf{D}) \propto
\textnormal{prob}(\textbf{D}|\textbf{X}) \times
\textnormal{prob}(\textbf{X}),
\end{equation}

\noindent where $\textnormal{prob}(\textbf{X}|\textbf{D})$ is the
probability of the model given the data (which is the quantity of
interest), $\textnormal{prob}(\textbf{D}|\textbf{X})$ is the
probability of the data given the model and
$\textnormal{prob}(\textbf{X})$ is the prior probability
distribution of the model parameters. Since we assume a flat prior
in all our model parameters above, \textnormal{prob}(\textbf{X}) =
constant so that

\begin{equation}
\textnormal{prob}(\textbf{X}|\textbf{D}) \propto
\textnormal{prob}(\textbf{D}|\textbf{X}).
\end{equation}

Assuming gaussian errors implies that

\begin{equation}
\textnormal{prob}(\textbf{D}|\textbf{X}) \propto \exp(-\chi^2/2),
\end{equation}

\noindent where $\exp(-\chi^2/2)$ is the likelihood function, with
$\chi^2$ defined in the standard way, as the sum of the normalized
residuals between the model-predicted observables $M_i$ and the
observed values $D_i$ i.e.

\begin{equation}
\chi^2 = \sum^N_{i=1} \Big ( \normalsize \frac{M_i-D_i}{\sigma_i}
\Big )^2 \normalsize
\end{equation}

The error that enters into the $\chi^2$ equation ($\sigma_i$) is
computed by adding, in quadrature, the observational uncertainties
with the errors adopted for the stellar models, which we assume to
be 0.05 mag in each optical filter \citep{Yi2003}.
$\textnormal{prob}(\textbf{X}|D)$ is a \emph{joint} probability
distribution, dependent on all the model parameters. From this
joint distribution, each free parameter is
marginalised\footnote{{\color{black}To isolate the effect of a
single parameter X1 in, for example, a two-parameter model
$[\textnormal{prob} (X | D) \equiv \textnormal{prob}(X1, X2 | D)]$
we can integrate out the effect of X2 to obtain the marginalized
probability distribution for X1:
$\textnormal{prob}(X_1|D)=\int^{\infty}_0
\textnormal{prob}(X_1,X_2|D) dX_2$.}} to extract its
one-dimensional probability distribution.} We take the median
value of this distribution as the best estimate of the parameter
in question. The 25th and 75th quartile values (which encompass
50\% of the probability) provide an estimate of the uncertainty in
the parameter. This yields, for every TD, a best estimate and
error for each free parameter. It is worth noting that the derived
error represents the combined uncertainty in the parameter
estimate due to the observational and model errors and the various
degeneracies between the free parameters.

\begin{figure}
$\begin{array}{c}
\includegraphics[width=3.5in]{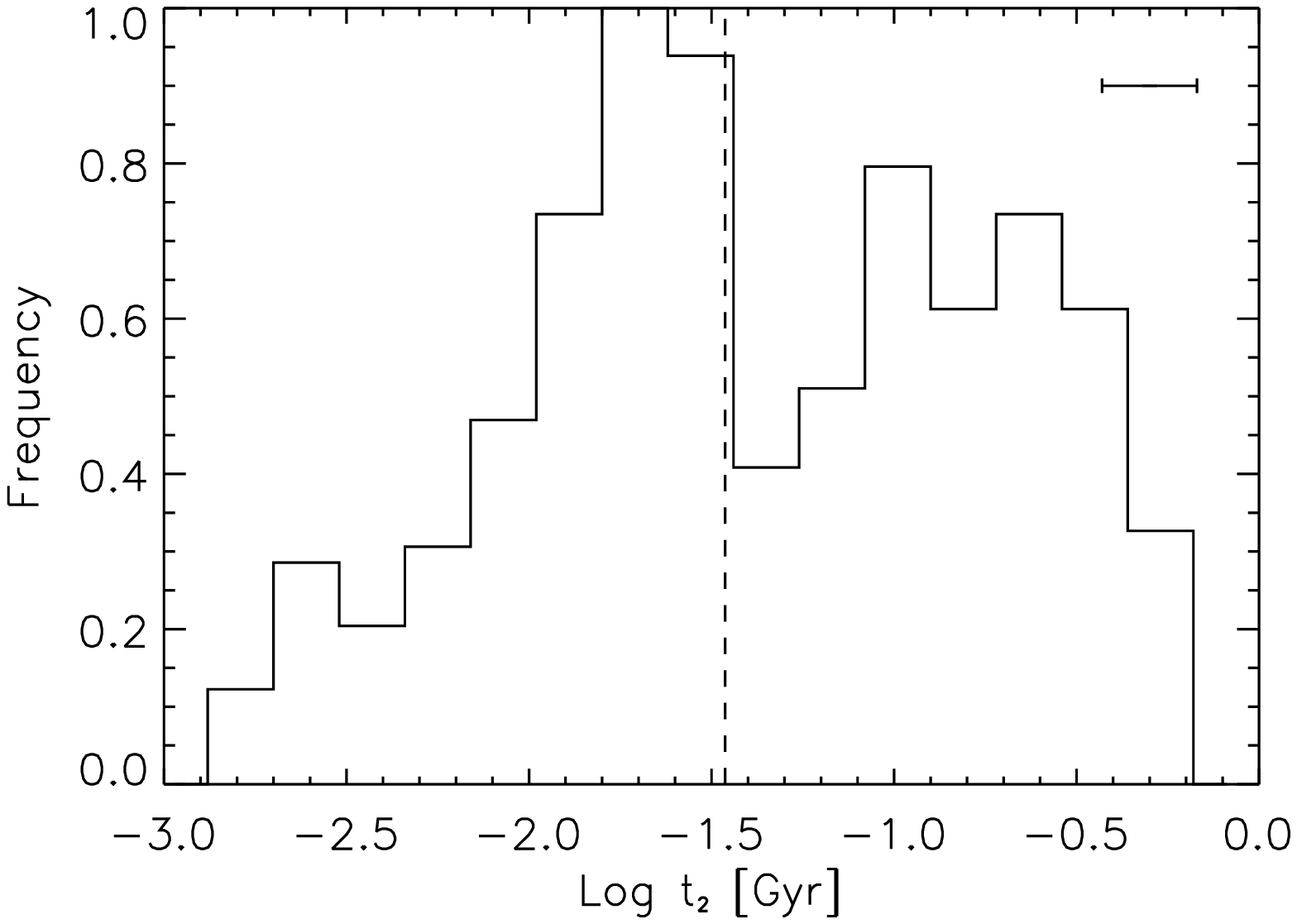}\\
\includegraphics[width=3.5in]{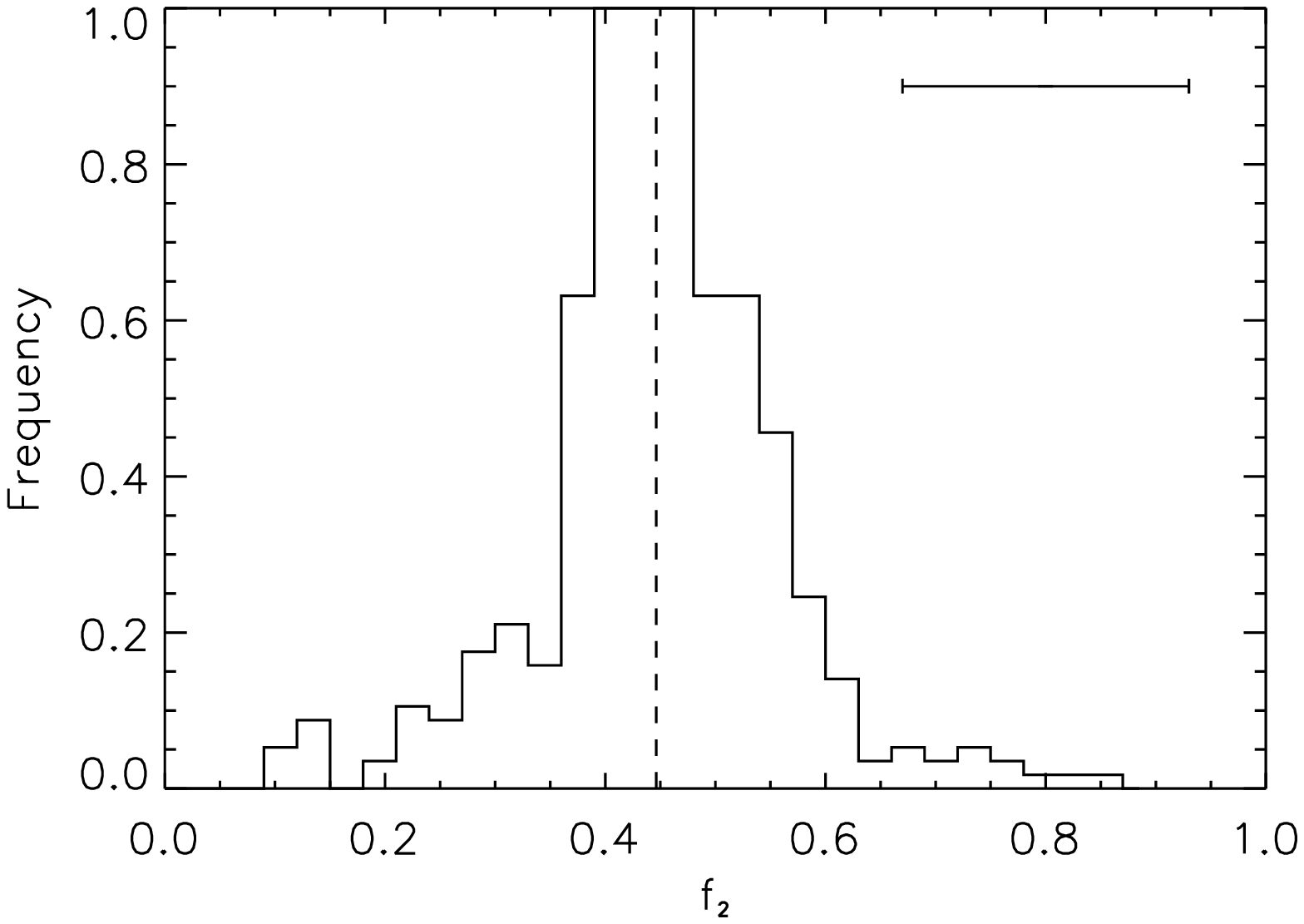}\\
\includegraphics[width=3.5in]{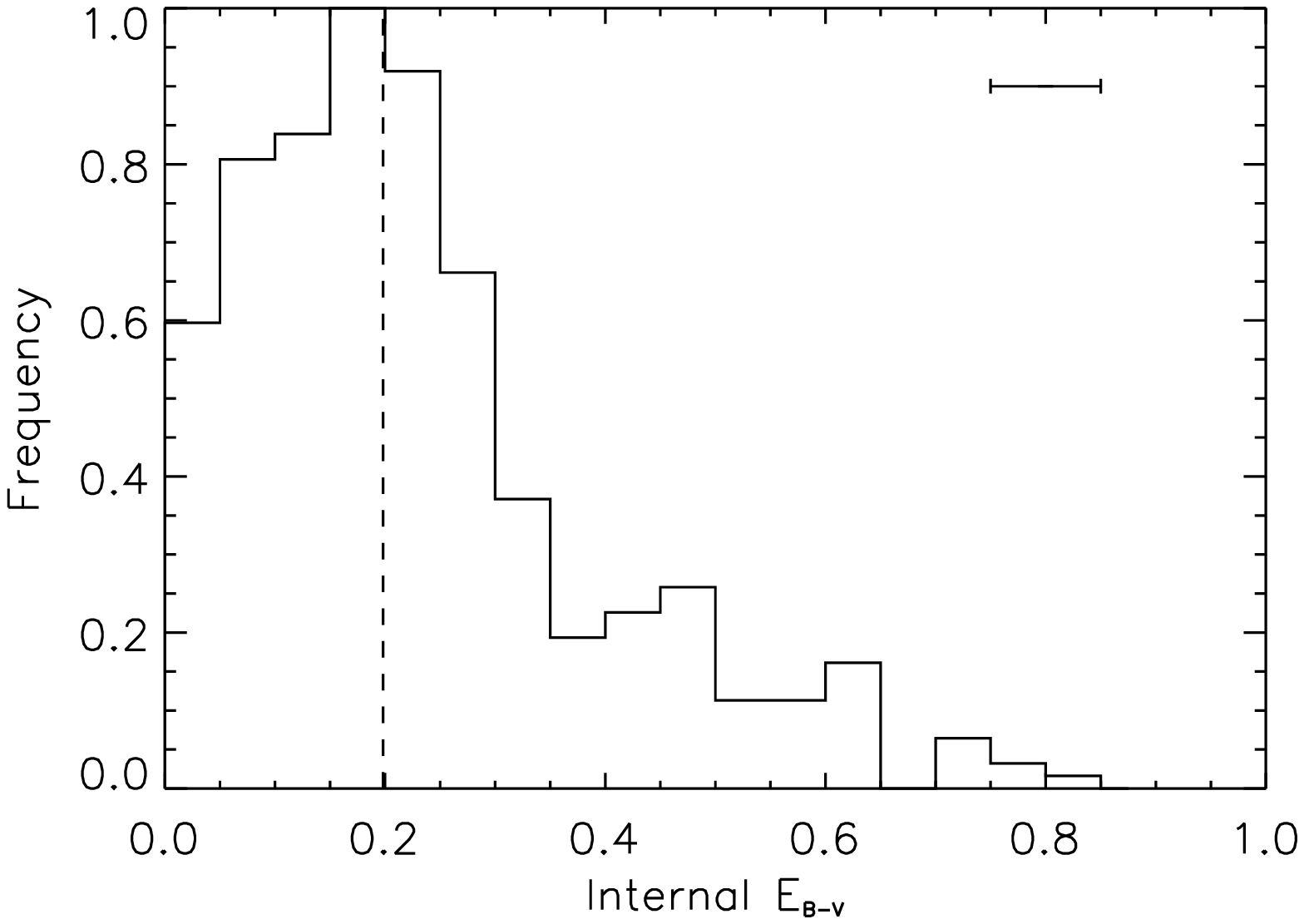}
\end{array}$
\caption{Estimated values of the free parameters that drive the TD
star formation histories: age ($t_2$; {\color{black}top}) and mass
fraction ($f_2$; {\color{black}middle}) of the young stars and
internal extinction in the galaxy ($E_{B-V}$;
{\color{black}bottom}). {\color{black}Median} values of the
distributions are shown using the {\color{black}dashed} lines.
While a substantial young stellar component exists in TDs, with
ages less than $\sim$0.5 Gyr (the median age is $\sim$30 Myr), an
equally significant component, drawn from old stars in the parent
disks, also appears to be present in these systems.}
\label{fig:td_sfhs}
\end{figure}

Figure \ref{fig:td_sfhs} presents the distribution of free
parameters for our TD population. Not unexpectedly, and in
agreement with the wider literature, we find that a substantial
young stellar component exists in the TDs, with ages less than
$\sim$0.5 Gyr (the median derived age is $\sim$30 Myr). The
derived mass fractions in young stars largely range between 20 and
80\% with a {\color{black}median} value of $\sim$45\%. The
estimated internal extinctions are typically lower than $E_{B-V}
\sim 0.5$ ($A_V \lesssim 1.5$), with a median value of $\sim$0.2.
These results indicate that a significant fraction (around half)
of the stellar content of TDs is likely to be composed of old
stars from the parent disks. While the mass-fraction uncertainties
are large (around $\pm$15\%), \emph{the bulk of the TDs are
inconsistent with a purely young stellar population}. It is
likely, therefore, that TDs are not formed purely through gas
condensations in the tidal tails but that their potential wells
contain significant contributions from pre-existing stellar
material from the parent disks.

%...................................................................................................

\section{Could the dwarf galaxy census have a significant
contribution from tidal dwarfs?} We conclude our analysis by
investigating the potential contribution of TDs to the dwarf
galaxy population observed in the Universe today. The analysis in
Section 3 indicates that TDs typically form in gas-rich or wet
major mergers that involve two spiral galaxies. If the number of
wet major mergers experienced by a massive galaxy over a Hubble
time is $N_{wet}$, the average number of TDs produced per wet
major merger is $N_{TD}$ and the fraction of TDs that survive for
a Hubble time is S, then the number of TDs expected per massive
galaxy today is estimated to be

\begin{equation}
N_{wet} \times N_{TD} \times S.
\end{equation}

Integration of the empirical major merger rate in massive galaxies
over time indicates that every massive galaxy typically
experiences $\sim 4$ major mergers over a Hubble time (Conselice
2007, see also Bell et al. 2006, Lotz et al. 2006). Typically, at
most one of these major mergers takes place after $z \sim 1$
\citep[e.g.][]{Conselice2003,Lin2004,Bell2006,Jogee2009}. Since
the merger activity at $z>1$ is likely to be dominated by
interactions between two gas-rich spiral galaxies
\citep[e.g.][]{Khochfar2003,Kaviraj2009}, this suggests that 3 out
of 4 of the major mergers experienced by a typical massive galaxy
are likely to be wet.

Theoretical work suggests that $\sim$50\% of TDs with \emph{masses
greater than $10^8$ M$_{\odot}$} are likely to survive for a
Hubble time \citep{Bournaud2010}. \emph{TD-producing} mergers each
create, on average, 1.2 TDs in this mass range. However, only
$\sim$18\% of major gas-rich mergers produce such TDs in the first
place. Hence, the average number of TDs with masses greater than
$10^8$ M$_{\odot}$ per gas-rich major merger is $\sim$0.22 (i.e.
1.2 $\times$ 0.18). Therefore the number of such TDs per massive
galaxy today is estimated to be $3 \times 0.22 \times 0.5 = 0.33$.

The \emph{observed} galaxy mass function indicates that dwarf
galaxies are the dominant galaxy type in the local Universe
\citep[e.g.][]{Sandage1985,vandenbergh1992,Sabatini2003}. The
ratio of dwarf to massive galaxies (D/M) in Coma
\citep{Secker1996}, restricted to dwarfs with masses greater than
$\sim 10^8$ M$_{\odot}$ ($M(r)<-14.5$), is $\sim$5.8. The
corresponding value in Virgo is very similar \citep{Ferguson1991}.
If the TD contribution to this ratio is $\sim$0.33 (as calculated
above) then $\sim$6\% of the dwarf population in clusters could
plausibly have a tidal origin.

It should be noted that this estimate assumes that the TD
production rate in high-redshift mergers is similar to that in
their nearby counterparts. Mergers at high redshift typically
involve higher gas masses \citep[e.g.][]{Daddi2010,Tacconi2010}
and may yield more TDs than their local counterparts
\citep[e.g.][]{Wetzstein2007}. However, simulations of
high-redshift major mergers \citep{Bournaud2011}, in which the
interstellar medium is more clumpy and turbulent than in their
nearby counterparts \citep[e.g.][]{Elmegreen2009}, suggest that
these interactions do not produce the long tidal tails seen in
local mergers. This may have implications for the lifetime of
tidal objects, since they are formed closer to their parent
galaxies, making them more vulnerable to disruption. Definitive
studies of the TD production rate at high redshift requires both
further simulation work and empirical studies of high-redshift
mergers at the peak epoch of stellar mass assembly \citep[$2<z<4$,
see e.g.][]{Madau1998,Hopkins2004,Hopkins2006} using
high-resolution data e.g. from the Wide Field Camera 3 (WFC3) or
the Extremely Large Telescopes (ELTs). Nevertheless, it is worth
noting that even if TD production rates were several factors
higher in the early Universe, it remains unlikely that the
\emph{entire} local dwarf galaxy census has a tidal origin.

%Note that restricting the massive TDs to those that form in tidal tips does not make much difference
%because they are almost all in the tidal tail tips.

%...................................................................................................

\section{Summary}
We have performed a statistical observational study of the TD
population in the local Universe, by exploiting a large,
homogeneous sample of galaxy mergers compiled from the SDSS DR6
using the Galaxy Zoo project. The aim of this work has been to
explore the statistical properties of local TDs, both to
complement existing observational studies (which are typically
based on relatively small samples of mergers) and as a comparison
to the wide body of theoretical work that has recently been
performed on the formation and evolution of TDs.

Our results indicate that 95\% of TD-producing mergers involve
interactions between two spiral galaxies, both typically residing
in the blue cloud. The overwhelming majority of these parent
systems have mass ratios greater than $\sim$1:7, reside in field
environments and are located within 15 optical half-light radii of
the parent galaxies. TD stellar masses are less than 10\% of the
stellar masses of their parents, with those forming at the tips of
tidal tails typically a factor of 4 more massive than those that
form at the base of the tails. TDs are typically bluer than their
parents, the median colour offset being $\sim$0.3 mag in $(g-r)$.
The presence of an AGN in the parent galaxies does not affect the
TD colours. It is worth noting that only around a fifth of
gas-rich major mergers produce massive TDs (with masses greater
than $10^8$ M$_{\odot}$).

An analysis of their star formation histories indicates that TDs
contain both newly formed stars and old stellar material drawn
from the disk of their parent galaxies. The young stellar
components have ages less than $\sim$0.5 Gyr, with a
{\color{black}median} derived age of $\sim$30 Myr in the TD
population as a whole. The young components contribute stellar
mass fractions between 20 and 80\%, with a typical value of
$\sim$45\%. The estimated internal extinctions are typically lower
than $E_{B-V} \sim 0.5$ ($A_V \lesssim 1.5$). The derived mass
fractions of young stars strongly suggest that TD formation is not
simply the result of gas condensations along tidal tails in
mergers. Stellar material from the parent disks contributes almost
equally to the mass in these objects.

Finally, we have explored the likely TD contribution to the dwarf
galaxy census in the nearby Universe. By combining the number of
gas-rich major mergers experienced by a massive galaxy over a
Hubble time with the average number of TDs expected to form in
each merger and their expected survival rate, we have estimated
the number of dwarfs per massive galaxy that are likely to come
from the TD population. Comparison to the observed ratio of dwarf
to massive galaxies in nearby clusters suggests that $\sim$6\% of
the dwarfs in local clusters may be of tidal origin, assuming that
the TD production rate in the nearby Universe is representative of
that in high-redshift mergers. Observational studies of TDs in
high-redshift mergers, using forthcoming data from the WFC3 and
the ELTs, are keenly anticipated to further explore the role of
mergers in the formation of the dwarf galaxy population at the
present day.

%...................................................................................................

\nocite{Gentile2007,Milgrom2007,Bell2006,Lotz2006,Conselice2007}

%...................................................................................................

\section*{Acknowledgements}
The anonymous referee is thanked for several constructive comments
on the manuscript. SK acknowledges fellowships from the Royal
Commission for the Exhibition of 1851, Imperial College London,
Worcester College, Oxford and support from the BIPAC institute at
Oxford. This publication has been made possible by the
participation of more than 250,000 volunteers in the Galaxy Zoo
project. Their contributions are individually acknowledged at
http://www.galaxyzoo.org/Volunteers.aspx. Support for the work of
KS was provided by NASA through Einstein Postdoctoral Fellowship
grant numbers PF9-00069, issued by the Chandra X-ray Observatory
Center, which is operated by the Smithsonian Astrophysical
Observatory for and on behalf of NASA under contract NAS8-03060.

Funding for the SDSS and SDSS-II has been provided by the Alfred
P. Sloan Foundation, the Participating Institutions, the National
Science Foundation, the U.S. Department of Energy, the National
Aeronautics and Space Administration, the Japanese Monbukagakusho,
the Max Planck Society, and the Higher Education Funding Council
for England. The SDSS Web Site is http://www.sdss.org/.

The SDSS is managed by the Astrophysical Research Consortium for
the Participating Institutions. The Participating Institutions are
the American Museum of Natural History, Astrophysical Institute
Potsdam, University of Basel, University of Cambridge, Case
Western Reserve University, University of Chicago, Drexel
University, Fermilab, the Institute for Advanced Study, the Japan
Participation Group, Johns Hopkins University, the Joint Institute
for Nuclear Astrophysics, the Kavli Institute for Particle
Astrophysics and Cosmology, the Korean Scientist Group, the
Chinese Academy of Sciences (LAMOST), Los Alamos National
Laboratory, the Max-Planck-Institute for Astronomy (MPIA), the
Max-Planck-Institute for Astrophysics (MPA), New Mexico State
University, Ohio State University, University of Pittsburgh,
University of Portsmouth, Princeton University, the United States
Naval Observatory, and the University of Washington.

%.........................................................................................................

\bibliographystyle{mn2e}
\bibliography{references}

%.........................................................................................................

\end{document}